# Topology-based Phase Identification of Bulk, Interface, and Confined Water using Edge-Conditioned Convolutional Graph Neural Network

A. Moradzadeh[1], H. Oliaei[1], and N. R. Aluru[2]

aluru@utexas.edu

[1]Department of Mechanical Science and Engineering, University of Illinois at Urbana−Champaign, Urbana, IL, 61801 United States, [2]Oden Institute for Computational Engineering and Sciences, Walker Department of Mechanical Engineering, The University of Texas at Austin, Austin, TX, 78712 United States

**Abstract**

Water plays a significant role in various physicochemical and biological processes. Understanding and identifying water phases in various systems such as bulk, interface, and confined water is crucial in improving and engineering state-of-the-art nano-devices. Various order parameters have been developed to distinguish water phases, including bond-order parameters, local structure index, and tetrahedral order parameters. These order parameters are often developed with the assumption of homogenous bulk systems, while most applications involve heterogeneous and non-bulk systems, thus, limiting their generalizability. Our study develops a methodology based on the graph neural network to distinguish water phases directly from data and to learn features instead of predefining them. We provide comparisons between baseline methods trained using conventional order-parameters as features, and a graph neural network model trained using radial distance and hydrogen-bonding information for phase classification and phase transition of water in bulk, interface, and confined systems with continuous and discontinuous phase transitions.

I.     **Introduction**

Water is an indispensable part of many physicochemical and biological phenomena, and its phase can significantly alter physicochemical and biological phenomena such as CO2 reduction,[1] proton transport,[2] power generation[3], and water desalination[4]. Furthermore, the properties of water such as diffusion coefficient, dielectric permittivity, and density as well as structural properties[5] depend on its phase. Therefore, it is of great importance to accurately identify different water phases. Like most other liquids, a significant understanding of water is developed through computational studies, where atomistic level data about the positions and velocities are available.[6,7] Therefore, the prediction of water phases from water molecules topology, *i.e.,* configuration of other molecules around a tagged water molecule, is a task worth studying and understanding, especially for confined systems such as carbon nanotubes (CNTs) due to CNTs technological applications.

Due to the high dimensionality and uninterpretable nature of atomistic simulation data, researchers have developed a wide variety of order parameters to reduce dimensionality and predict the phase of a system from reduced dimensions. Motivated by the importance of water in various areas, water is studied through multiple order parameters (OPs) such as bond-order parameter (BOP),[8,9] tetrahedral order parameter[10], and local-structure index.[11,12] Even though these order parameters are widely adopted in various studies ranging from ice nucleation[13], phase discrimination/identification,[8] liquid-liquid transitions,[14,15] free energy calculation[16], they are far from ideal. In many cases, it requires a lot of domain expertise and effort to combine multiple order parameters to reach conclusive findings or even define new order parameters.[16,17] The problem is particularly pronounced for confined systems as OPs are usually defined for homogenous and bulk systems, which is not the case for confined water. Furthermore, due to the interplay between fluid-fluid and fluid-wall interactions in the confined system,[18] confined systems have richer physics accompanied by anomalous behavior in the phase transition region, where both continuous and discontinuous phase transition can occur (the discontinuous phase transition characterized by a sharp change in the potential energy, enthalpy, or OP of the system, while continuous phase transition shows only a critical point).[19,20] Various computational and experimental studies are performed to investigate and identify phase behavior of confined water;[21,22] however, many challenges remain and one of them is to predict the phase of water directly from positional information, especially for the confined system.

Similar to the order parameter design in the phase identification task is the design of kernel and feature engineering in image, speech, and text processing applications, which required lots of domain expertise and human time.[23–25] During the last few decades, however, the process of kernel and feature engineering has been revolutionized by deep learning-based methods, which are adapted for a wide variety of applications in physics, chemistry, and biology. Water, as one of the most complex and important liquids, has been successfully studied using various deep learning methods in applications such as force field development and phase-identification.[26–34] However, recent deep-learning methods still try to use traditional OPs as features for the phase-identification of water, which does not address the issue of order parameter definition for non-homogenous systems.

The main bottleneck of phase identification stems from the nature of the data obtained from MD simulations. The data used for training of machine learning models should ensure that input features be permutation, rotation, and translation invariant.[23] The atomic coordinates obtained from MD simulation do not possess these properties, which hinders the application of many conventional deep learning algorithms unless some sort of transformation is applied. Initial attempts for classification of water phase using deep learning-based methods started by a study where multiple features requiring multiple transformations are fed into a multilayer perceptron. However, the method requires arbitrary rotational transformations of the dataset to enforce the rotational invariance. Recent progress in graph neural networks (GNNs) provides a suitable tool to deal with the atomistic data as they are best described in a non-Elucidation space.[35] In addition to addressing permutation, rotation, and translation invariance, GNN addresses the variable size of the data, which is the case for confined water with different number of neighbors depending on water phase and its distance from the wall. GCIceNet[36] is developed to solve the problem of rotational and permutational invariance using GNN. Even though GCIceNet is successful, GCIceNet constructs node features using OPs, which is an edge feature (it depends on the distance between atoms). Additionally, OPs used as node features are not well-defined for confined systems.

In this study, we use the latest advances in GNN, particularly edge-conditioned convolutional (ECC) graph neural network, to address the problem of phase identification of water in bulk, interface, and confined systems in an end-to-end fashion.[37] In short, ECC is successfully applied to the point Cloud dataset, which mimics the problem of phase identification in many ways. We formulate the phase-identification problem as a graph classification task and use the ECC layers to remove the need for human-engineered order parameters. To do so, we construct our graphs $G = \{V, E\}$ by collecting the oxygen atoms within the cut-off distance of a tagged oxygen atom; based on the performance and computational cost, we keep all the oxygens or several closest oxygen atoms. The oxygen atoms form the nodes ($V$) of the graph, and the pairwise distance between all oxygen atoms (nodes) is the edge feature. The node feature ($X \in \mathbb{R}^{|V|\times 2}$) is the one-hot encoded vector $\{0,1\}$, where the tagged oxygen atom *i.e.*, the water molecule whose phase we want to predict, has a different node feature compared to its neighboring oxygen atoms. We use a full adjacency matrix to classify phases ($A = [1]_{|V|\times|V|} - I_{|V|}$). We collect all the pairwise distances between all nodes as the edge feature of our graphs. Additionally, we collect information regarding the hydrogen bonds between water molecules by determining whether an edge corresponds to a donor-acceptor or acceptor-donor hydrogen-bond as well as no-hydrogen-bond. The hydrogen bonding is the only information incorporated as a feature in neural network training, as nodes corresponding to it are coarse-grained in graph representation. The dimension of an edge, therefore, is a vector of size 4 ($E \in \mathbb{R}^{|V|\times|V|\times 4}$). The output of the graph classification task for bulk water is a vector of dimension $n_c$, where $n_c$ is the number of different water phases in the dataset. We study 9 different phases of water such as Ih, Ic, II, III, VI, VII, VIII, and IX Ices as well as liquid water. For interface and confined systems, we use similar input as for a bulk system, but the output is a binary value indicating whether water is liquid or solid. GNNs are trained to predict whether a particular configuration of atoms is liquid-like or solid-like inside CNTs or at interface for various temperatures. We study CNT 10x10, inside which both continuous and discontinuous phase transition can occur. Our reference solid and liquid systems used for training are picked from temperatures away from phase transition temperature. The model successfully shows both sharp and smooth changes in the fraction of liquid-like molecules near the phase transition, allowing us to predict phase transition temperature faster than normal methods. This is an advantage of our method as previous studies need averaging over lots of trajectories to calculate OP or thermodynamic properties to obtain phase transition temperatures.

The rest of the paper is organized as follows. First, we describe the details of MD simulation and calculation of order parameters, followed by training of graph neural network and random forest models and comparison between their performance. Finally, we summarize the findings of our study.

**II. Method**

**MD Simulations:**

Molecular dynamics (MD) simulations of water in bulk and confinement are performed using the GROMACS package.[38] Water is modeled using TIP4P/Ice model as it performs better for phase transitions.[39] For water at interface, we study Ice h/vapor interface, where a quasi-liquid layer[40] can form at the interface of solid and vapor due to the missing hydrogen-bonding in the interface of solid and vapor. For confined cases, carbon-water interactions are modeled using the parameters from reference[41]. The temperature and pressure of the systems are controlled using the Nosè-Hoover thermostat with a time constant of 0.2 ps and Parrinello-Rahman barostat with a time constant of 2.0 ps, respectively.[42] Initial configurations of the bulk system are generated using GenIce package.[43] After energy minimization steps on the initial bulk configurations, MD simulation is performed for 25 ns at corresponding temperature and pressure of phase (see SI for temperature and pressure of each phase). The data for machine learning model training as well as OPs are obtained from the last 10 ns of simulation. For the confined systems, we fill CNTs using a reservoir. Once filled, the isolated periodic CNT mimicking an infinite CNT is simulated at different temperatures, by gradually decreasing temperature from 390 $K$ to 10 $K$ with the rate of 1 $K/ns$. For every 10 $K$ decrease, we simulate the system for 20 ns, again the last 10 ns are used for post-processing.

To compare the performance of GNN with conventional machine learning methods as a baseline, we calculate OPs including local-structure index (LSI), BOP, and tetrahedral OP. The LSI indicates the translational order of the system, and it considers $|\mathcal{N}(i, r_{cf} = 0.37\ nm)|$ neighboring water molecules by ordering them in ascending pairwise distances $(r_{j+1} > r_j\ \forall\ j \in \mathcal{N}(i, r_{cf} = 0.37\ nm))$. Mathematically, it is defined as,

$$LSI = \frac{1}{|\mathcal{N}(i)|} \sum_{j \in \mathcal{N}(i)} [\Delta(j) - \bar{\Delta}]^2 \tag{1}$$

where $\Delta(j)$ is the difference between the pairwise distance of two neighboring water molecules *i.e.* $(\Delta(j) = r_{j+1} - r_j)$ and $\bar{\Delta}$ is the average value of $\Delta(j)$.

The BOP of order $l$ $(q_l)$ is the other OP used in the baseline machine learning method, where it is a coarse-grained representation of Steinhardt parameter $q_{lm}$,[8,9] which can be expressed as follows,

$$q_{lm}(i) = \frac{1}{|\mathcal{N}(i,\ r_{cf} = r_{cf,6})|} \sum_{j \in \mathcal{N}(i)} Y_{lm}(\theta_{ij}, \phi_{ij}) \tag{2}$$

where $Y_{lm}$ is the spherical harmonic function of degree $l$ and order $m$. $\theta_{ij}$ and $\phi_{ij}$ are polar angles. The cut-off distance $(r_{cf} = r_{cf,6})$ of the neighbor list is chosen such that $|\mathcal{N}(i,\ r_{cf} = r_{cf,6})|$ equals 6. The BOP of order $l$ and degree $m$ is defined as,

$$Q_{lm}(i) = \frac{1}{|\mathcal{N}(i,\ r_{cf} = r_{cf,6})| + 1} \left( q_{lm}(i) + \sum_{j \in \mathcal{N}(i)} q_{lm}(\theta_{ij}, \phi_{ij}) \right) \tag{3}$$

which is coarse-grained by averaging over degree through the following expression,

$$q_l(i) = \sqrt{\frac{4\pi}{2l+1} \sum_{m=-l}^{l} |Q_{lm}|^2} \tag{4}$$

The tetrahedral OP is defined based on the four nearest molecules and takes a value between 0 and 1, where 0 and 1 correspond to an ideal gas and perfect tetrahedral respectively. It can be expressed as,

$$q_{te} = 1 - \frac{3}{8}\sum_{j=1}^{3}\sum_{k=j+1}^{4}(\cos\psi_{jk} + \frac{1}{3}) \tag{5}$$

where $\psi_{jk}$ is the angle formed from the tagged molecule and two of the four closest water molecules. BOPs and tetrahedral OP calculated using PyBoo package are used to train the baseline machine learning models.[44] In this study, we use the random forest as our baseline.[45] We also store the pairwise distance between atoms within a cut-off distance from a tagged water molecule as an edge feature for GNN training along with the presence and type of hydrogen-bonding as a one-hot-encoded vector.

**Machine Learning**

To investigate the performance of GNN over other machine learning algorithms, we train a baseline machine learning method *i.e.,* random forest (RF). Before going into the details of our training, we describe the task and procedure we have taken. We treat the problem as a classification problem, and for the bulk system, our output is a one-hot encoded vector of 9 different phases. The task in the confined and interface systems, however, is simplified to the identification of solid and liquid phases, where the output is binary *i.e.,* 0 or 1. The data for classification is selected from multiple temperatures, both above and below the melting temperature of TIP4P/Ice models, exposing the model to both liquid and solid phases. The performance of the method is of special interest for confined systems, where determination of the phase transition temperature from the conventional method (first-order change in order parameters) is not straightforward and requires a long simulation. It can also suffer from significant noise as the timescale of phase transition can be large for continuous phase transition. However, the model can still predict phase behavior well (more quantitative analyses are provided later).

### Random Forest

Random forest algorithm (RF) as an ensemble learning method selects a subset of features (in our cases a vector of dimension 7, composed of LSI and tetrahedral order parameters as well as BOPs with degrees of {4,6,8,10,12}. For each selected subset, a decision tree is trained by randomly selecting a subset of features and dataset, followed by the construction of decision trees. Once the training of various trees is done, a majority vote is taken to determine the class for a given data. We apply grid search with 5-fold cross-validation to obtain optimal depth and number of trees for RF.

### Graph Neural Network

The graphs in this study are denoted by $G = \{X, E\}$, where $X = \{x_i \in \mathbb{R}^2 | i = 1, ..., N\}$ is the set of all nodes (oxygen atom of water molecules) with a dimension of 2 corresponding to one-hot-encoding based on whether the node is a tagged oxygen atom or a neighboring oxygen atom, and $E = \{e_{ij} \in \mathbb{R}^4 | x_i, x_j \in X\}$ is the set of edges with 4 attributes *i.e.,* the pairwise distance and the one-hot-encoding of hydrogen-bonding corresponding to donor-acceptor, acceptor-donor, or no-hydrogen-bond cases. The graph can also be described using the adjacency matrix, a binary matrix of dimension $|X|^2$. Elements of the adjacency matrix determine connections between nodes $A = [\delta_{ij}]_{|X| \times |X|}$. Note that depending on the computational cost and classification performance, we use different numbers of water molecules to form the graph (see Figure 1 for schematic representations of graph construction).

In general, most of the GNN methods belong to the message-passing networks, which utilize combinations of message, aggregation, and update.[35] In this study, we use the ECC layer to build our GNN model.[37] The hidden representation of nodes $h^l$ at layer $l$ is equal to a weighted sum of hidden representation $h^{l-1}$ in its neighborhood. The weights in the ECC are generated by another network, also known as filter generating networks, which is usually modeled with a MLP with trainable parameters. Mathematically, the following operations are performed in the *l*-th layer,

$$h^l(i) = h^{l-1}(i)W_{root}^l + \sum_{j \in N(i)} h^{l-1}(j)\text{MLP}(e_{ji}, w^l) + b^l \qquad (6)$$

where $b^l$ is the $l$-th layer bias, and $N(i)$ is the neighborhood of node $i$ ($N(i) = \{j; (j,i) \in E\}$). $h^l$ are embedding of the $l$-th layer. Note that $h^0$ corresponds to input feature $x$ and $w^l$ are learnable parameters of the multi-layer perceptron *i.e.* weight generating function (MLP above). $W_{root}^l$ are learnable weights corresponding to contribution from the hidden representation of $i$-th node itself. After 3 or 4 ECC layers, we use a pooling function (sum pooling) to find a representation for each graph. The role of pooling is to reduce node embeddings of the whole graph into a single vector. Additionally, the pooling layer should be invariant to the permutation of nodes, we use sum function as our pooling layer. The pooled representation is fed to a multi-layer perceptron with 1 hidden layer. All the layers, except the last of multi-layer perceptron, use the ReLU activation function. Further details regarding the structure of layers are given in the SI.

To train the parameters of the GNN we use either binary- or categorical-cross-entropy losses defined as,

$$\mathcal{L} = -\sum_{i \in n_c} p_i \log v_i \qquad (7)$$

where $p_i$ and $\log v_i$ are $i$-th element of vectors with dimension equal to the number of classes ($n_c$) representing one-hot-encoding and GNN predictions, respectively. The Adam optimizer is used to train the model with a learning rate of 0.00005 for 100000 epochs with early stopping if the loss is saturated for 5 consecutive epochs on the validations dataset (0.2 of the dataset). The batch size of 16, 32, or 64 is used depending on the computational cost. Spektral and TensorFlow packages[46,47] are used to perform GNN training, and the top2phase package developed as a python package for broader usage (see GitHub link for the package).

**III. Results and Discussions**

We first perform data-exploratory analysis to examine the sufficiency of BOPs to separate different bulk water phases (bulk systems are shown in Figure 2a). To do so, we obtain 2D scatter plots of BOPs for two sets of degrees, namely ($q_6, q_8$) and ($q_{10}, q_{12}$). In Figure 2b, we show the results of the analysis, where we observe a large overlap between any selected BOPs. Following this step, we train RF methods with different numbers of trees and depth and find the optimal RF parameters. The model reaches an average accuracy of 89.2 percent. Training the GNN model with the same dataset leads to an average accuracy of 99.9 percent. Figure 3 shows more quantitative analysis on the accuracy of both GNN and RF models based on the confusion matrix. The confusion matrix shows the percentage of dataset misclassified for off-diagonal elements, and diagonal elements show the percentage of correctly classified samples per class. The GNN confusion matrix (Figure 3b) shows far superior behavior as off-diagonal elements are far less than their counterparts in the RF confusion matrix (Figure 3b).

Along the same lines, we also study Ice h/vapor system, where we simulate the system for the temperature range of 10 K to 300 K with a 10 K step (see Figure 4 for Exploratory data analysis as well as the schematic representation of waters at different temperatures before and during phase transition). The experimental and computational investigation show formation of a quasi-liquid layer at the interface of Ice h /vapor.[40,48–50] The predicted melting temperature from the experiment and simulation is around $270 \pm 5$ K. We select temperatures of [10,140] K and [290, 300] K as our reference solid and liquid systems for ML training. After training the model, which shows an accuracy of 99% for GNN and 0.97% for RF, we feed the data from other temperatures to predict melting temperature and compare classification results obtained from RF and GNN. To do so, we compare the liquid fraction at each temperature using both RF and GNN models in Figure 5. Additionally, we show the potential energy of the system, which shows a sharp change near melting temperature. The GNN predicts a melting temperature of 275 K and RF predict it as 275 K, and potential energy indicates a melting temperature of $275.0 \pm 2.5$ K. The behavior of GNN is monotonic within the temperature range of our study, showing an increasing number of liquid-like molecules, while RF shows a non-monotonic and inconsistent behavior with temperature increase.

Next, we study confined systems, where we simulate water confined inside (10,10) CNT for the temperature range of 10 K to 390 K with a 10 K step (see Figure 6 for Exploratory data analysis as well as the schematic representation of waters in liquid and solid phase of water with average densities of 16.75 $nm^{-3}$ and 19.14 $nm^{-3}$). Confined water in general shows more complex behavior compared to bulk water; for example, with the increase in density, phase transition becomes continuous, especially for CNTs with a smaller diameter. This phenomenon is usually attributed to the interplay between interface-water and water-water interactions. As shown in figure 6a-d, the solid phase of water inside (10,10) CNT show heptagon and heptagon with single-file water, respectively at densities of 16.75 $nm^{-3}$ (low) and 19.14 $nm^{-3}$ (high), while liquid phases of both densities inside CNT looks like each other. The larger overlap of scatter plots of BOPs in high-density cases shown in Figure 6e-h indicates difficulty in using BOPs. Note that the high-density case corresponds to a continuous phase transition. The predicted melting temperatures for low- and high-density cases are around $270 \pm 10$ K and $290 \pm 10$, respectively. The trend is consistent with previous computational studies. Like the Ice h/vapor case, we select two representative temperatures for both liquid and solid – in this case, we use [10,150] K and [310, 380] K as our reference solid and liquid systems, respectively. After training the model, the GNN model achieves 0.994 and 0.949 accuracy, respectively for low- and high-density cases; the accuracy of RF is lower than that of GNN with GNN and RF achieving 0.997 and 0.809 accuracies, respectively for low and high densities (see SI for confusion matrix and more details on model performance). The GNN outperforms RF in the high-density case, thereby demonstrating the capability of GNN in complex cases. The lower accuracy of RF is attributed to the large overlap of BOPs as shown in Figure 6 e-h. GNN, however, learns its featurization based on data and does not face many difficulties in distinguishing solid and liquid phases. Once the models are trained, we feed the data from other temperatures to predict melting temperature and compare the results obtained from RF and GNN. In Figure 7a-b, we compare the liquid fraction at each temperature using both RF and GNN models for low- and high-density cases. Additionally, we show the potential energy and axial diffusion coefficient, which exhibit a sharp change near melting temperature for low density and smooth transition for high density, signatures of discontinuous and continuous phase transition. The predicted melting temperatures are close to MD simulation results using both RF and GNN models. However, the behavior of GNN is more monotonic with temperature change as shown in Figure 7a-b. For example, the fractions of liquid-like molecules are a non-decreasing function of temperature, while RF shows a non-monotonic and inconsistent behavior with temperature increase. We also note that the larger deviation of both GNN and RF for the high-density case can be attributed to the difficulty in reaching complete equilibrium in the high density as the timescale of relaxation of simulation is large. Overall, the results of the high-density case prove the abilities of GNN in more complex environments.

## IV. Conclusions

In this study, we train a graph neural network model to classify different phases of water in bulk, interfacial, and confined environments. To address the issue with the definition of order parameters in the confined environments, we train the model to learn features from the positional data, *i.e.,* the distance between the oxygen atom of tagged molecules and all other water molecules oxygen atoms within a cut-off distance. We augmented the edges features with the hydrogen-boding information (acceptor-donor, donor-acceptor, or lack of hydrogen-bonding) as hydrogen atoms are coarse-grained in the graph representation. The results show successful employment of model in bulk, interfacial, and confined water inside a carbon nanotube, especially in terms of its generalization compared to baseline method trained using classical order parameters model. Furthermore, the predicted melting temperature and behavior of the model in both continuous and discontinuous phase transition inside carbon nanotube were in good agreement with the change in the potential energy and dynamics of waters. In summary, the methodology presented here provides a robust data-driven tool to classify and study the phase behavior of complex systems.

**Acknowledgments**

This work was supported by the National Science Foundation under Grant 2140225. The authors acknowledge the use of Blue Waters supercomputing resources at the University of Illinois at Urbana-Champaign. Furthermore, this work partially used the Extreme Science and Engineering Discovery Environment (XSEDE) Stampede2 at the Texas Advanced Computing Center through allocation TG-CDA100010. This work also utilized resources supported by the National Science Foundation's Major Research Instrumentation program, grant #1725729, as well as by the University of Illinois at Urbana-Champaign.

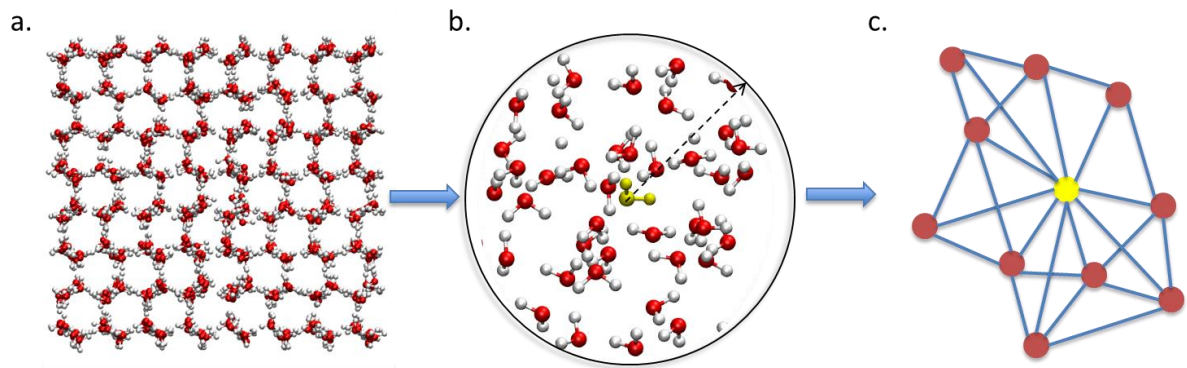

Figure 1. Schematic representation of water and corresponding graph structure representing the structure. a. atomistic configuration b. neighbor list formation based on a tagged water molecular c. graph representation with nodes as oxygen atoms, and edges representing connection with blue color. Each edge has four dimensions, representing distance and H-bond. Node color represents whether it is the tagged molecule or not.

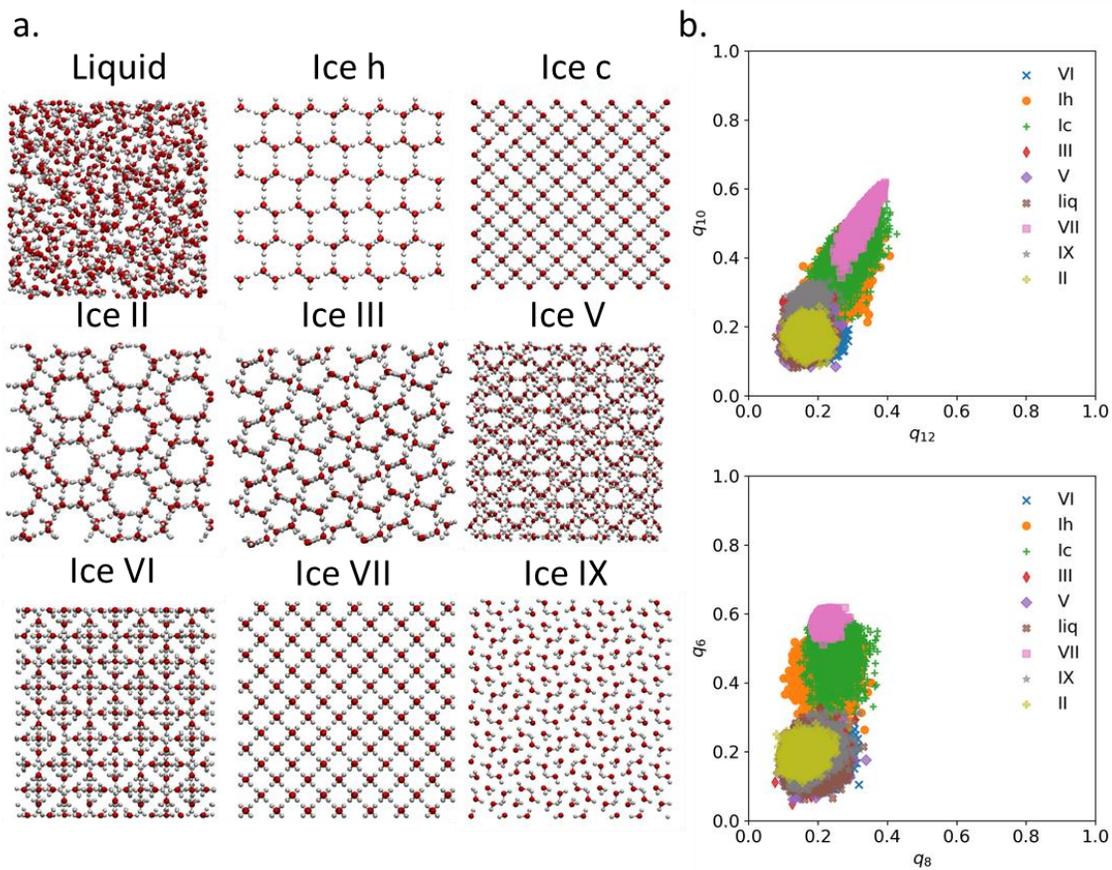

Figure 2. Exploratory data analysis of bulk water phases. a. Schematic representation of different bulk water phases b. scatter plot of $(q_{10}, q_{12})$ and $(q_6, q_8)$ of bulk phases of water.

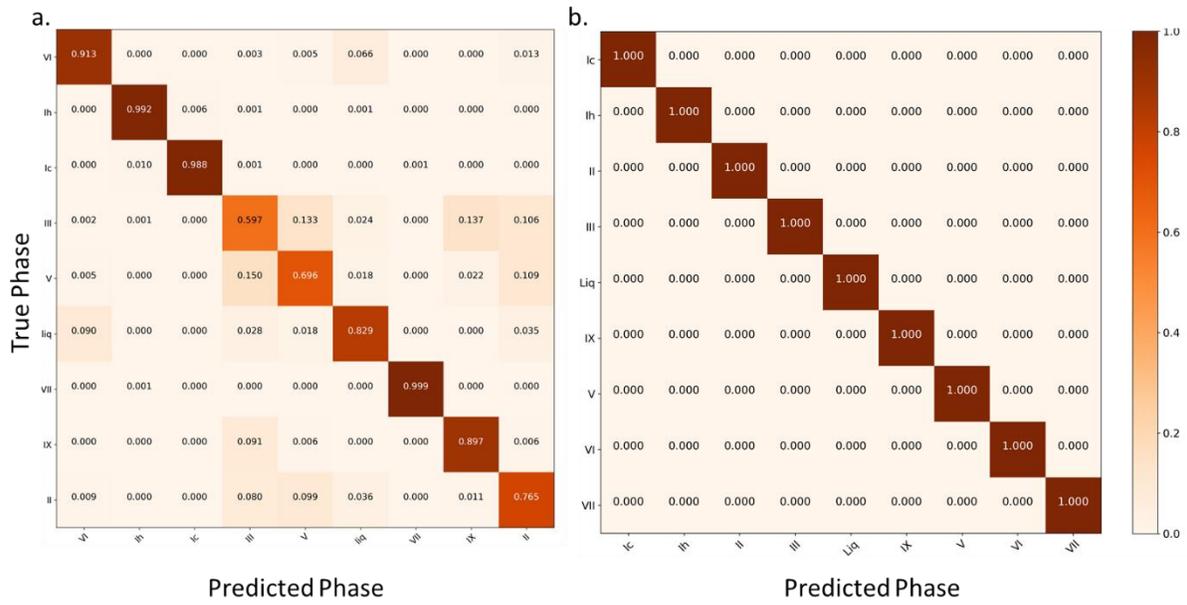

Figure 3. Confusion matrix for classification of bulk water. a. using RF. b. using GNN.

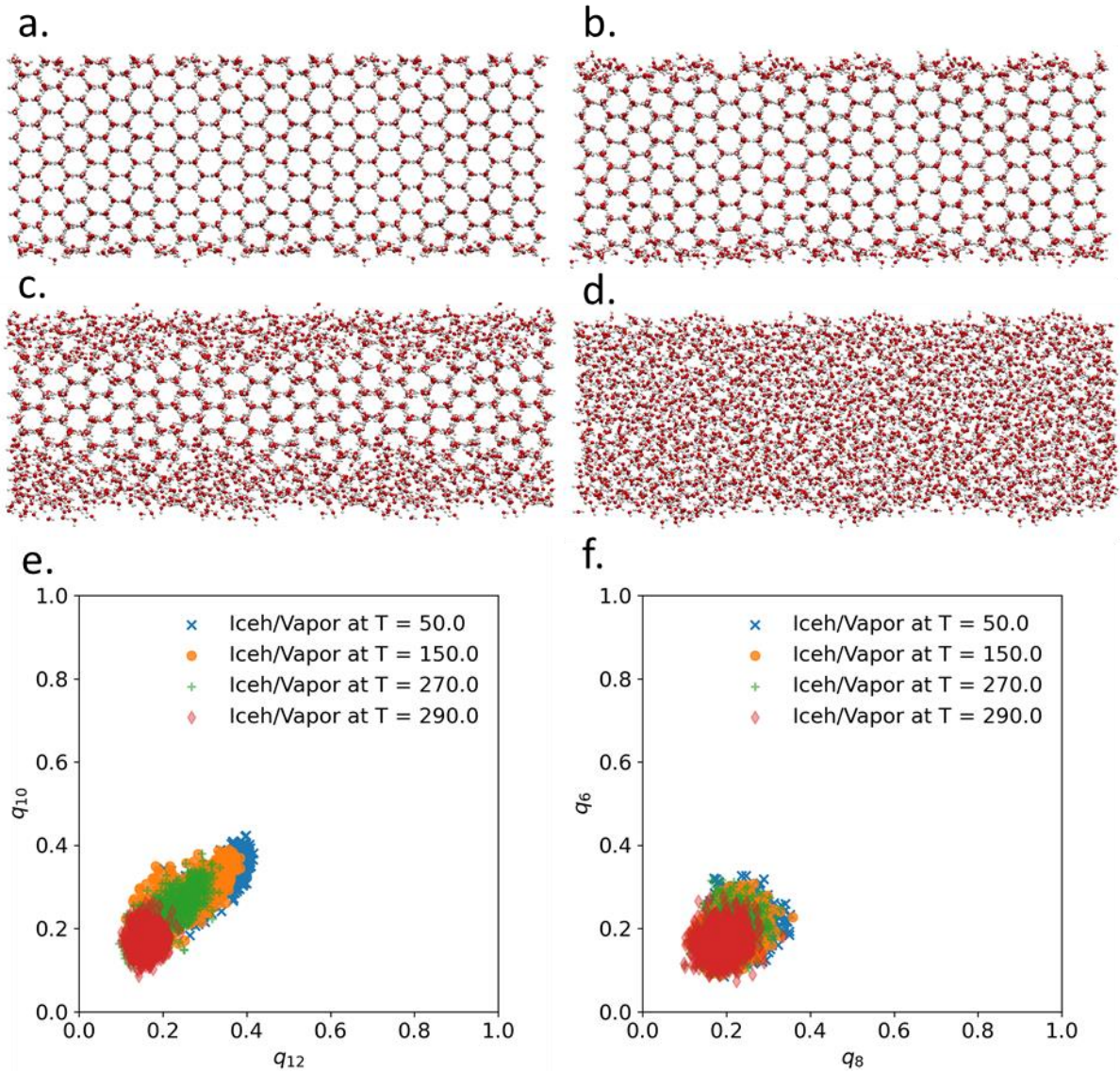

Figure 4. Schematic representation of Ice h/vapor interface at different temperatures along with exploratory data of analysis. a. configuration at 50 K b. configuration at 150 K. c. configuration at 270 K d. configuration at 290 K e. scatter plot of $(q_{10}, q_{12})$ at different temperatures f. scatter plot of $(q_6, q_8)$ at different temperatures. Distribution of order parameter pairs shows significant overlap at different temperatures.

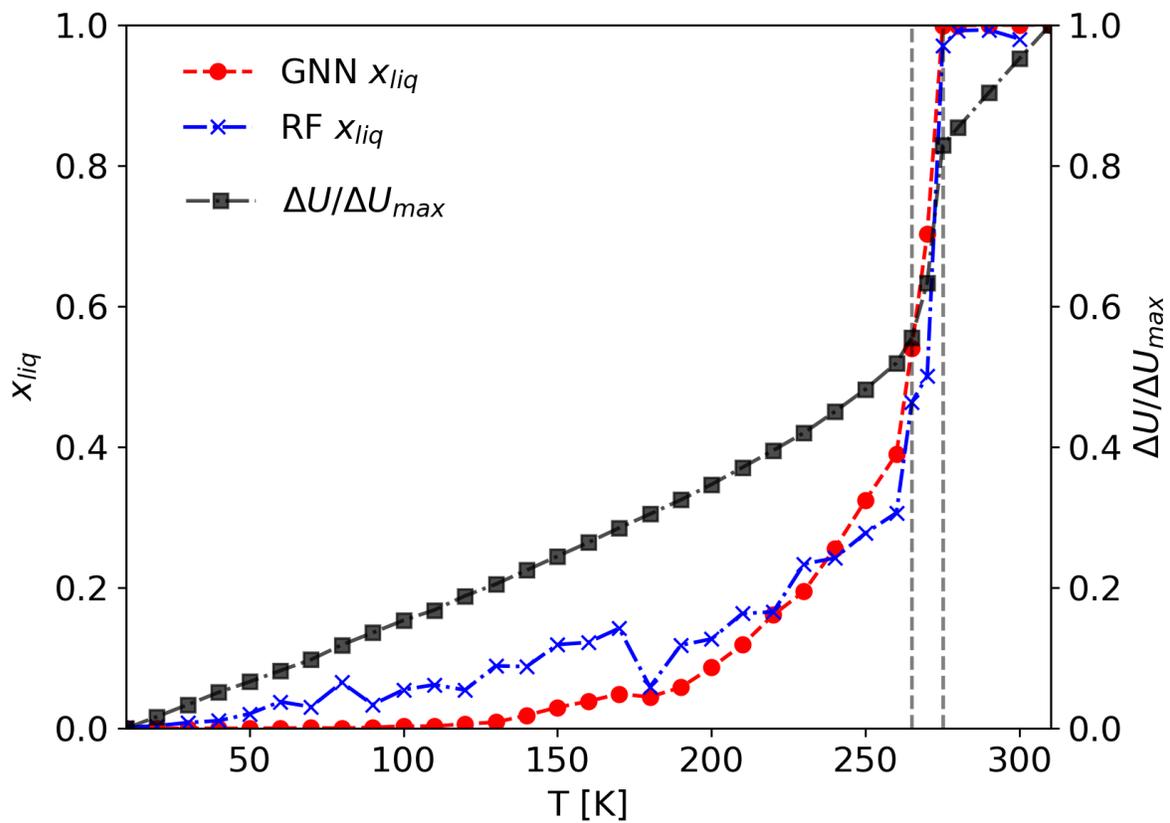

Figure 5. Phase transition of Ice h/vapor system. The black line with squares shows scaled potential energy of water at different temperatures. The red circle and blue cross represent fraction of liquid-like molecules at different temperatures obtained using RF and GNN, respectively. Dashed vertical lines indicate temperature range at which all Ice-like molecules disappear due to temperature increase.

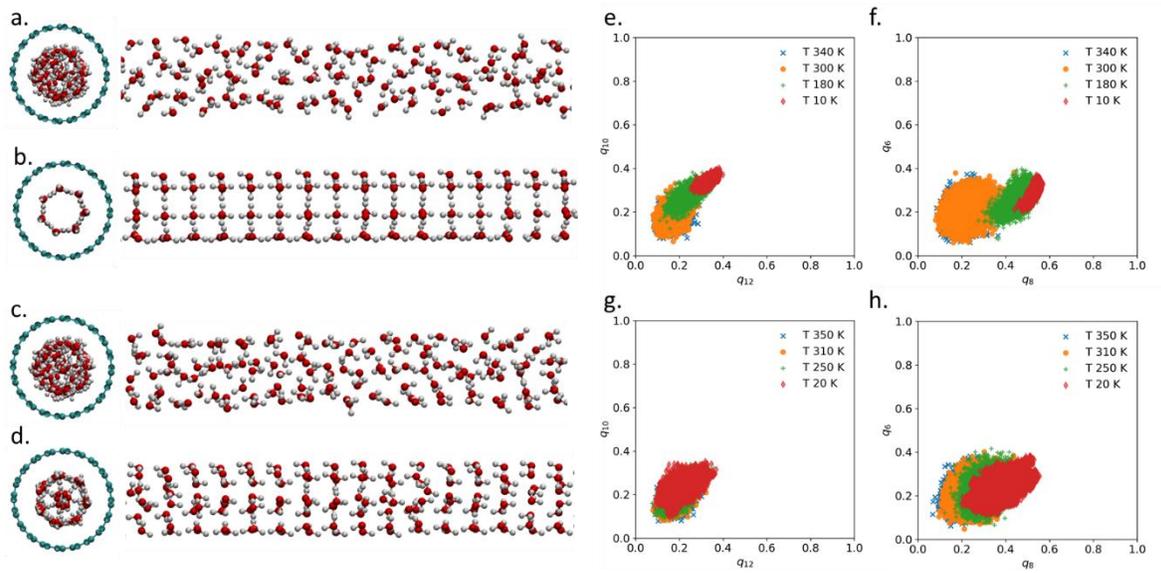

Figure 6. Schematic representation of confined water inside a (10,10) CNT at different temperatures and densities along with Exploratory data of analysis. a. liquid water configuration at 350 K and 16.75 $nm^{-3}$. b. solid water configuration at 20 K and 16.75 $nm^{-3}$. c. liquid water configuration at 350 K and 19.14 $nm^{-3}$. d. solid water configuration at 20 K and 19.14 $nm^{-3}$. e. scatter plot of $(q_{10}, q_{12})$ at different temperatures and density of 16.75 $nm^{-3}$. f. scatter plot of $(q_6, q_8)$ at different temperatures and density of 16.75 $nm^{-3}$. g. scatter plot of $(q_{10}, q_{12})$ at different temperatures and density of 19.14 $nm^{-3}$. h. scatter plot of $(q_6, q_8)$ at different temperatures and density of 19.14 $nm^{-3}$. Distribution of order parameter pairs show significant overlap for different phases.

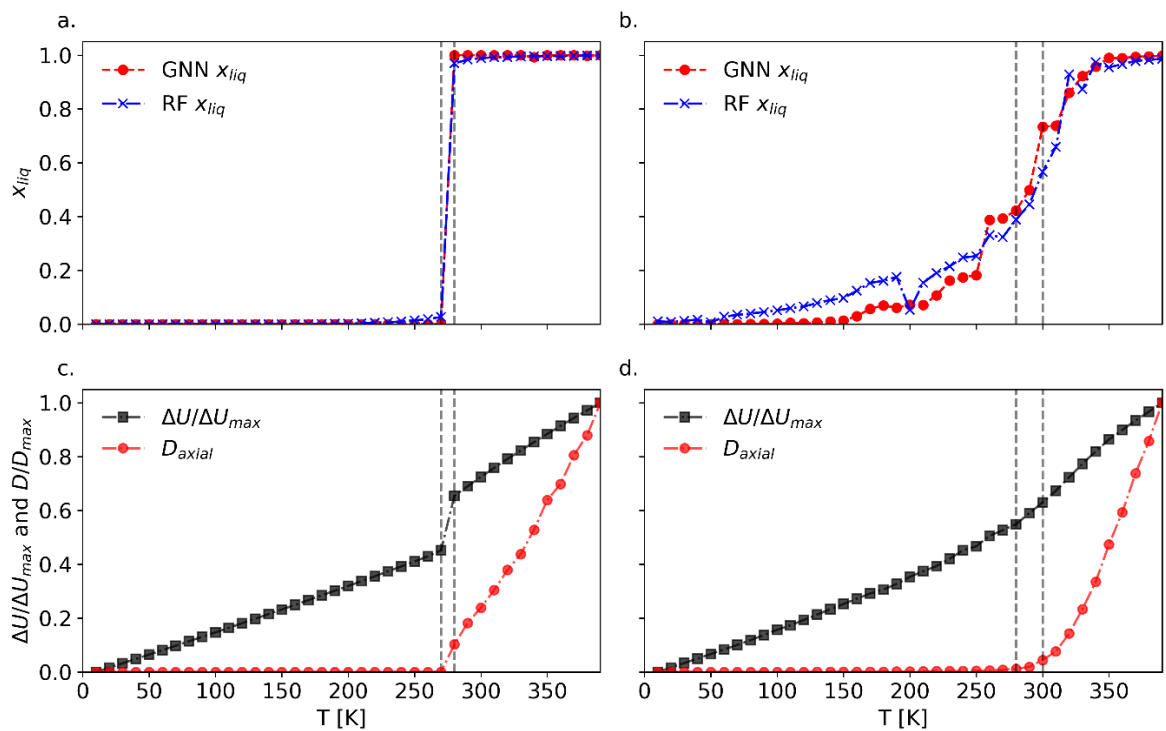

Figure 7. Phase transition of confined water with discontinuous and continuous phase transition. Comparison between the normalized potential energy and diffusion coefficient change at different temperatures and two densities. a. fraction of liquid-like molecules for CNT with density of 16.75 $nm^{-3}$ predicted using RF and GNN models. b. fraction of liquid-like molecules for CNT with density of 19.14 $nm^{-3}$ predicted using RF and GNN models. c. potential energy and diffusion coefficient at different temperatures for CNT with density of 16.75 $nm^{-3}$. d. potential energy and diffusion coefficient at different temperatures for CNT with density of 19.14 $nm^{-3}$. vertical lines show phase transition region. In a-b, red circles and blue crosses show results of RF and GNN models, respectively. In c-d, potential energy and diffusion coefficients are show with black squares and red circles, respectively.

# Supporting Information:

# Topology-based Phase Identification of Bulk, Interface, and Confined Water using Edge-Conditioned Convolutional Graph Neural Network

A. Moradzadeh[1], H. Oliaei[1], and N. R. Aluru[2]

aluru@utexas.edu

[1]Department of Mechanical Science and Engineering, University of Illinois at Urbana−Champaign, Urbana, IL, 61801 United States, [2]Oden Institute for Computational Engineering and Sciences, Walker Department of Mechanical Engineering, The University of Texas at Austin, Austin, TX, 78712 United States

# 1. Refence Systems

Simulations of following systems are performed to obtain data for training of neural networks.

| Phase Name | $T\ [K]$ | $P\ [bar]$ |
|---|---|---|
| Liquid | 300 | 1.0 |
| Ih | 240 | 1.0 |
| Ic | 150 | 1.0 |
| II | 200 | 5,000 |
| III | 253 | 3,000 |
| V | 253 | 6,000 |
| VI | 253 | 10,000 |
| VII | 350 | 100,000 |
| IX | 50 | 3,000 |

# 2. Confusion Matrix of the CNT

Comparision between the confusion matrix of RF and GNN models, indicating better performance of GNN for confined water system. Figure S1 shows the confusion matrix for classification of solid and liquid phase the inside CNT filled with water at the average density of 16.75 $[\#/nm^{-3}]$. Both models perform fairly well in this density with slide better performance for GNN. In Figure S2, we show the confusion matrix for classification of solid and liquid phases inside the CNT filled with water at the average density of 19.14 $[\#/nm^{-3}]$. The high density case is most challenging for both model due to the continuous phase transition. The GNN does a superior classification compared to the RF with 5 to 15 % higher accuracy for different phases.

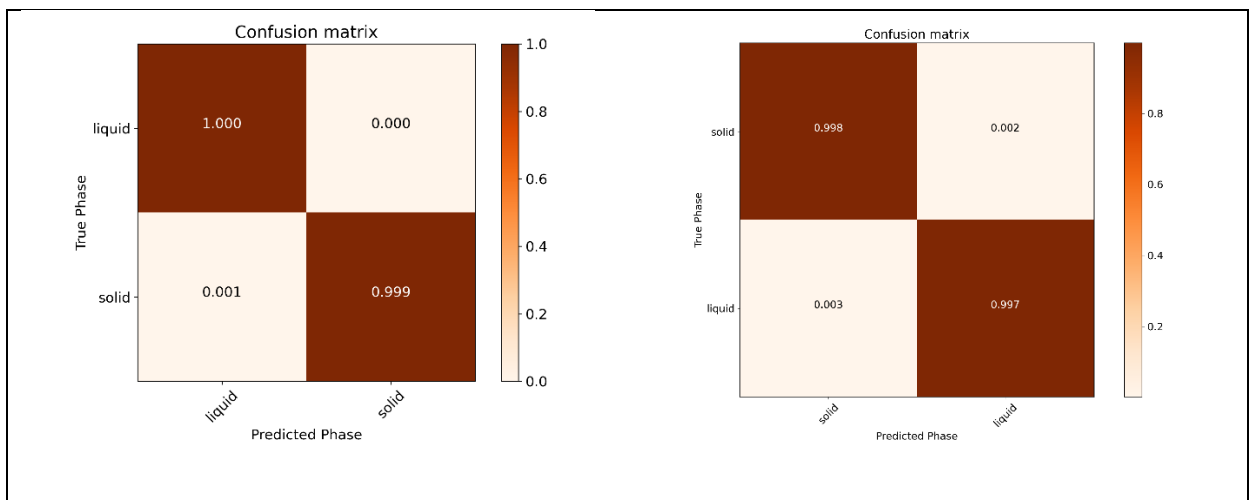

Figure S1. Confusion matrix for solid and liquid phase using GNN (left) and RF (right). Water confied inside CNT with average density of 16.75 $[\#/nm^{-3}]$

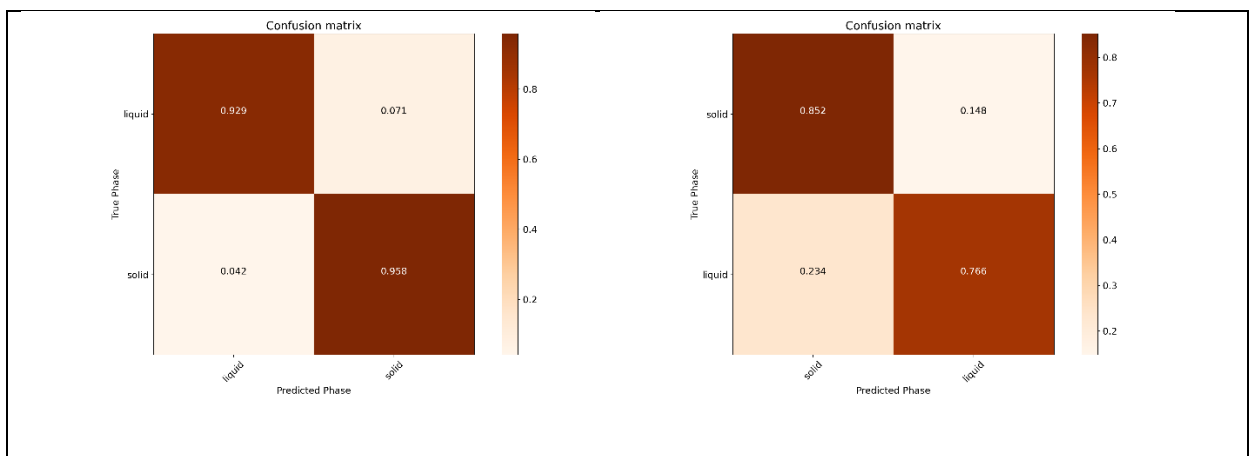

Figure S2. Confusion matrix for solid and liquid phase using GNN (left) and RF (right). Water confied inside CNT with average density of 19.14 $[\#/nm^{-3}]$